\documentclass[]{interact}

\usepackage{epstopdf}
\usepackage[caption=false]{subfig}

\usepackage[numbers,sort&compress]{natbib}
\bibpunct[, ]{[}{]}{,}{n}{,}{,}
\renewcommand\bibfont{\fontsize{10}{12}\selectfont}
\makeatletter
\def\NAT@def@citea{\def\@citea{\NAT@separator}}
\makeatother

\theoremstyle{plain}

\theoremstyle{definition}

\theoremstyle{remark}

\usepackage{natbib,bibentry}

\begin{document}


\title{Quick and Simple Kernel Differential Equation Regression Estimators  for Data with Sparse Design}

\author{
\name{Chunlei Ge\textsuperscript{a}\thanks{CONTACT Chunlei Ge. Email: chunlei.ge@ubc.ca} and W. John Braun\textsuperscript{b}}
\affil{\textsuperscript{a}UBC Okanagan; \textsuperscript{b}UBC Okanagan}
}

\maketitle

\begin{abstract}
Local polynomial regression of order at least one often performs poorly in regions of sparse data.  Local constant
 regression is exceptional in this regard, though it is the least accurate method in general, especially at the boundaries
 of the data.  Incorporating information from differential equations which may approximately or exactly hold is one
 way of extending the sparse design capacity of local constant regression while reducing bias and variance.   A nonparametric regression method that exploits first order differential equations is introduced in this paper and applied to noisy mouse tumour growth data. Asymptotic biases and variances of kernel estimators using Taylor polynomials with different degrees are discussed. Model comparison is performed for different estimators through simulation studies under various scenarios which simulate exponential-type growth. 
\end{abstract}


 \section{Introduction}

Modelling data where there are relationships between outcome variables and possible explanatory variables remains an active area of research. Generalized linear models as a group of typically parametric regression models are popular when they can clearly explain and interpret how predictors influence the outcome. Compared with nonparametric regression, parametric regressions are more interpretable but less flexible and accurate due to constraints that they impose on the shape of the functional relationships between the response and explanatory variables. These need to be predetermined. In practical problems, scatterplots, which are not appropriately fit by straight lines and other parametric curves, can be analyzed by nonparametric methods, a proposed approach which adapts to the functional relationship and highlights unexpected features of the data.

Local polynomial regression, a nonparametric technique for smoothing scatter plots and modelling functions, was introduced by \cite{nadaraya1964estimating} and by \cite{watson1964smooth}. Typically, these regression models can reduce the noise or variablity in the data by averaging values within a certain range. The tools used in the fitted model are weighted polynomials inside the local neighborhood with the bandwidth of $h$, which is a smoothing parameter. The books by \cite{fan1996local} and \cite{loader2006local} provide several methods of bandwidth selection. The ability of capturing local nuances in data confers some valuable advantages on local polynomial regression models such as adaptive smoothing, local interpretability, and no need for explicit function specification. However, the dependence of local neighborhoods can lead to the difficulties in regions of sparse data. The information from differential equations can be a great assistance for local polynomial regression models.

The approach discussed in this paper is one specific form of differential equation assisted local polynomial regression. With the assistance of differential equations, local polynomial regression can capture more complex patterns in the data, for example, data with a sparse design. The improved model can corporate with different kinds of differential equations including linear and nonlinear forms. In this paper, we will introduce a relatively simple model, a differential equation assisted local exponential growth model.

\subsection{Kernel Differential Equation Regression Model}

We consider kernel regression models with the assistance of first order differential equations.

Given $n$ independent observations on an explanatory variable $x$ and
a response variable $y$,   
we consider models of the form 
\[y_i = g(x_i) + \varepsilon_i \mbox{ where } g'(x) = F(x,g(x)), \ \ \ \ i = 1, 2, \ldots, n,  \]
for some Lipschitz continuous function $F$ and uncorrelated mean-zero errors $\varepsilon_i$.   
We assume that the design points have been randomly sampled from an interval $[a, b]$
according to a probability density function $f$, or they have been selected according to
a fixed sampling design in that interval.  

\subsection{Differential Equation Assisted Local Exponential Growth Model}

We start a regression analysis using the above model with a specific form, $F(x,g(x))=\lambda g(x)$. Then the model will be an exponential growth model

\begin{equation}
y_i = g(x_i) + \varepsilon_i \mbox{ where } g'(x) = \lambda g(x), \ \ \ \ i = 1, 2, \ldots, n.  
\label{equ:exp} 
\end{equation}

We will call it local exponential growth model in this paper, or in short, local growth model. An application example will be performed in the next section. The data we used is mouse tumor growth data with sparse design.

\subsection{An Application Example of Sparse Tumor Growth Data }

At various points in this paper, we will consider the following example which  concerns
a set of control data for a chemotherapy trial in an animal experiment. 
Tumour volume measurements were taken in a single mouse.   The full set of observations  
is below.   Times are in days, and volumes are in cubic centimetres.  

 \begin{table}[ht]
\centering
\scalebox{1}{
\begin{tabular}{ccc}
  \hline
  & time & volume \\ 
  \hline
1  &  21  & 0.05 \\
2  &  25  & 0.09 \\
3  &  28  & 0.22 \\
4  &  31  & 0.32 \\
5  &  33  & 0.61 \\
6  &  35  & 0.70 \\
7  &  38  & 0.90 \\
8  &  40  & 1.29 \\
9  &  42  & 1.77 \\
10 &  45 &  3.32 \\
  \hline  
\end{tabular}
}
\caption{The full set of observations in mouse tumor volume data} 
\label{table:fullmouse}
\end{table}

\begin{figure}
        \centering
        \vspace{-18mm}
        \includegraphics[width=1\textwidth]{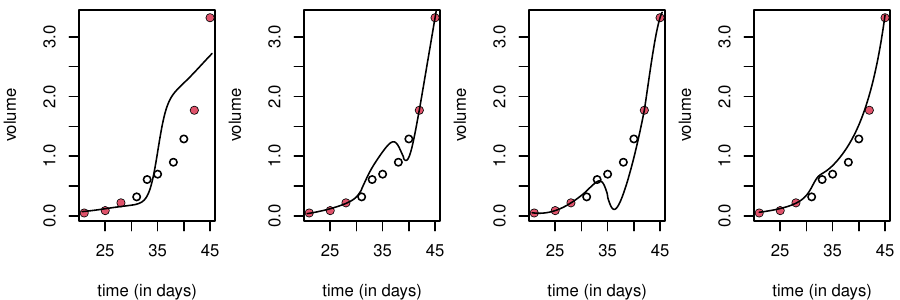}
        \vspace{-10mm}
        \caption{Mouse tumour growth data with some artificially missing observations (shown as open circles). Left panel: local constant model;   centre left panel: local linear model;   centre right panel: local quadratic model; right panel: local growth model.  All models are fit to the observed data (filled circles) using a gaussian kernel with bandwidth $h = 3.5$.}\label{fig:sparse}
        \end{figure}

We artificially remove some of the data to illustrate performance of various local polynomial models on sparse data.  The resulting data set is as follows and is plotted as the filled circles in Figure  \ref{fig:sparse}.  The removed data are denoted by open circles in that figure.  

\begin{table}[ht]
\centering
\scalebox{1}{
\begin{tabular}{ccc}
  \hline
  & time & volume \\ 
  \hline
1  &  21  & 0.05 \\
2  &  25  & 0.09 \\
3  &  28  & 0.22 \\
4  &  42  & 1.77 \\
5 &  45 &  3.32 \\
  \hline  
\end{tabular}
}
\caption{The mouse tumor volume data in sparse design after removing 5 observations.} 
\label{table:sparsemouse}
\end{table}

Local constant regression can sometimes handle this kind of data sparseness, but
not accurately, especially near boundaries.  Higher order local polynomial methods fail.   In 
particular, we fit local constant, local linear and local quadratic models to the second data 
set using a gaussian kernel with bandwidth $h = 3.5$ in each case.   As can be seen
in the first three plots, each method has deficiencies, caused by the gap in the data set.  Either
spurious bumps are introduced (as in the local linear and quadratic cases) or gross inaccuracy
results (in the local constant case).  

Our objective is to exploit the low variance and data sparseness-handling properties of 
local constant regression while achieving accuracy of higher order local polynomial methods. 
To do this, we will use additional information (or beliefs) about the regression function.  
Specifically, we suppose that the regression function is the solution of a differential equation. 
For example, a simplistic model for tumour volume is the exponential growth model:
\[ g' = \lambda g, \]
for some positive constant $\lambda$.  The explicit solution to this differential equation is 
\[ g(x) = g(0) e^{\lambda x}.   \]
This model could be fit directly to the data, upon estimating $\lambda$ by applying
least-squares estimation to the loglinear version of the model.  

Alternatively, we can obtain a  local solution of the differential equation by 
constructing a Taylor expansion of $g(x_i)$ about an evaluation point $x_0$, adapting the approach
described in detail by \cite{fan1996local}.  The first order approximation is 
\begin{equation}\label{eqn:Taylor} g(x_i) = g(x_0) + (x_i-x_0) g'(x_0) + O((x_i-x_0)^2). 
\end{equation}
Substituting the information from the differential equation, 
\[ g'(x_0) = \lambda g(x_0), \]
we have 
\[ g(x_i) = g(x_0) + (x_i-x_0) \lambda g(x_0) + O((x_i-x_0)^2). \]
There is only one local parameter to estimate: $g(x_0)$.  This can be estimated from
\[ \sum_{i=1}^n \left(y_i - g(x_0)(1 + (x_i - x_0) \lambda)\right)^2K_h(x_i - x_0) \]
where $K_h$ is a symmetric probability density function (the kernel) having scale $h$ (the
bandwidth).

   The plot in the right panel of Figure \ref{fig:sparse} shows the result of this approach
   for a grid of evaluation points $x_0$ in the interval $[21, 45]$, with bandwidth simply chosen
   as one-half of the median of the successive differences in the sorted covariate values.  

The result is not perfect, since there is a small spurious bump in the curve in the region
of the deleted data points. However, in comparison with the other curves in the figure, it
appears that the differential equation has provided useful information.   In a later
section of this paper, we return to this example with a somewhat more rigorous analysis.  

An important reference in the area of estimation of differential equation parameters and fitting
differential equations to data is \cite{ramsay2007parameter}.  The approach taken there is based on splines and requires extensive tuning. \cite{liang2008parameter} have proposed a different kernel-based differential equation estimation approach based on a measurement error model which is similar yet distinct from the method proposed in the present paper.  Their paper contains a large number of references to the literature on the fitting of ordinary differential equations to data.  
  
The rest of this paper proceeds as follows.  In Section 2, we outline our proposed
method.  The next section contains the theoretical demonstration, for the local exponential growth model, that the bias is reduced compared with local constant regression while the variance is not greatly increased. Section 4 features a simulation study showing how the method is
competitive with standard local polynomial approaches.  We also compare the
method with a nonlinear least-squares estimation to the solution of a differential
equation.  Comparisons are made both when the solution corresponds to the correct
model and when truth departs from the differential equation solution.  We also
study the effectiveness of our approach for a sparse design by simulation.  
The next section treats the mouse tumour growth data in detail by modifying the exponential growth model.  The paper concludes
with a short discussion in Section 6.

\section{Proposed Method}

Our general approach is to adapt 
local polynomial regression as described by  \cite{fan1996local}. 
For a given evaluation point $x_0$, a differential equation-assisted local polynomial regression estimator of $g(x_0)$ can be obtained
by minimizing the local least-squares objective function
\[ \sum_{i=1}^n (y_i - g^*(x_i))^2 K_h(x_i-x_0). \]
Expanding $g(x_i)$ in Taylor series up to degree $k$ about $x_0$ gives  
\[ g(x_i) \doteq g(x_0) + (x_i-x_0) g'(x_0) + \cdots + \frac{(x_i-x_0)^k}{k!} g^{(k)}(x_0).  \]
We therefore define
\[ g_k^*(x_i) = g(x_0) + (x_i-x_0) g'(x_0) + \cdots + \frac{(x_i-x_0)^k}{k!} g^{(k)}(x_0).  \]
Using the differential equation information, we observe that
\[ g'(x_0) = F(x_0, g) \]
and 
\[ g''(x_0) = F_1(x_0, g) + F_2(x_0, g) g'(x_0) = F_1(x_0, g) + F_2(x_0, g)F(x_0, g), \]
where $F_j$ denotes the partial derivative of $F$ with respect to its $j$
argument.   Additional derivatives can be computed, using symbolic algebra software if necessary.

Although there may be unknown global parameters governing the differential equation, 
there is  only one local parameter to estimate: $g(x_0)$.

\section{Theoretical Properties}

\subsection{The Proposed DE-Assisted Estimators}

We suppose the data follow the exponential growth model
\[ y_i = g(x_i) + \varepsilon_i,  \ \ \ \ \ x_i \in [a,b], \ \  i = 1,...n\]
where
\[ g' = \lambda g \]
and where we assume
$\lambda$ is a known parameter. We further assume conditions on the design density, 
kernel, and bandwidth as in \citet{fan1996local}.

Applying the first-degree Taylor expansion only:
\[
\sum_{i=1}^n (y_i - g(x_i))^2K_h(x_i - x) \] \[\doteq 
\sum_{i=i}^n (y_i - g(x) - g'(x)(x_i - x))^2 K_h(x_i - x) \]
\[ = \sum_{i=i}^n (y_i - g(x) - \lambda g(x) (x_i - x))^2 K_h(x_i-x) \]
leads to the DE-Assisted first-degree estimator
\[ \widehat{g}_1(x) = \frac{
\sum_{i=1}^n y_i(1+(x_i-x)\lambda)K_h(x_i-x)}{\sum_{i=1}^n (1 + (x_i-x)\lambda)^2 K_h(x_i-x)}.\]

where the subscript 1 indicates applying the first-degree Taylor expansion. We call it the local DE1-1 estimator, which is obtained with the known information of first order differential equation.

Similarly, we consider the second-degree estimator. Applying the second-degree Taylor expansion for $g(x_i)$ in a sufficiently small neighborhood of $x$, we have
\[
\sum_{i=1}^n (y_i - g(x_i))^2K_h(x_i - x) \] \[\doteq 
\sum_{i=1}^n (y_i - g(x) - g'(x)(x_i - x) - \frac{1}{2}g''(x)(x_i - x)^2)^2 K_h(x_i - x) \]
\[ = \sum_{i=1}^n (y_i - g(x) - \lambda g(x) (x_i - x) -  \frac{1}{2}\lambda^2 g(x)(x_i - x)^2)^2 K_h(x_i-x) \]
which leads to the DE-Assisted second-degree estimator
\[ \widehat{g}_2(x) = \frac{
\sum_{i=1}^n y_i(1+(x_i-x)\lambda + \frac{1}{2} (x_j -x)^2\lambda^2)K_h(x_i - x)}{\sum_{i=1}^n (1 + (x_i-x)\lambda^2)^2 K_h(x_i-x)}.\]

where the subscript 2 indicates applying the second-order Taylor expansion. We call it the local DE1-2 estimator.

Generally, we can find the general form of $k^{th}$-degree Estimator by applying the $k^{th}$-degree Taylor expansion we mentioned in section 2

\begin{align}
\widehat{g}_k(x) 
&= \arg\min_i \sum_{i=1}^n \left\{y_i- \sum_{p=0}^k \frac{1}{p!}(x_i-x)^p g^{(p)}(x)\right\}^2 K_h(x_i-x) \nonumber\\
&= \arg\min_i \sum_{i=1}^n \left\{y_i-\sum_{p=0}^k \frac{1}{p!}(x_i-x)^p \lambda^{p}g(x)\right\}^2 K_h(x_i-x) \nonumber\\
&= \frac{\sum_{i=1}^n \{y_i \sum_{p=0}^k \frac{1}{p!}(x_i-x)^p \lambda^{p} \} K_h(x_i-x) }{\sum_{i=1}^n \{\sum_{p=0}^k \frac{1}{p!}(x_i-x)^p \lambda^{p} \}^2K_h(x_i-x)}
\end{align}

\noindent where $g^{(p)}(x)=\lambda^p g(x)$, $p=1,2,...,k$.

We call it the local DE1-k estimator, where 1 indicates first-order differential equation and k indicates $k^{th}$-degree Taylor expansion.

In the next two sections, we will compute the conditional asymptotic biases and variances of $\hat{g}_k (x)$.

When we perform the conditional asymptotic analysis of the $k^{th}$ degree estimators , we make the following assumptions for our model (1)

\begin{itemize}
  \item $g(x)$, the mean function, has a bounded and continuous $k+1^{th}$ derivative in a neighborhood of $x$. 
  \item $f(x)$, the design density, is twice continuously differentiable and positive.
  \item $K(\cdot)$, the kernel function, is a symmetric and bounded PDF with finite sixth moment and is supported on a compact interval, say $[a,b]$.  
  \item \[R(K)=\int K^2(w)dw < \infty \] for the kernel function $K(\cdot)$.     
\end{itemize}

We will discuss the asymptotic conditional bias and variance of the DE-assisted estimators in the interior of $[a,b]$.

\subsection{The Asymptotic Conditional Bias}  

We compute the asymptotic conditional bias of the $k^{th}$-degree estimator $\hat{g}_k (x)$

\begin{equation}
\begin{split}
\mathrm{Bias}(\widehat{g}_k (x)|x_1,...,x_n) &=\mathrm{E}[\hat{g}_k(x)|x_1,...x_n]-g(x) \\
& \approx \frac{1}{(k+1)!}\lambda^{k+1} g(x) \frac{\sum_{i=1}^n (x_i-x)^{k+1}\{\sum_{p=0}^k \frac{1}{p!}(x_i-x)^p \lambda^{p}\}K_h(x_i-x) }{\sum_{i=1}^n\{\sum_{p=0}^k \frac{1}{p!}(x_i-x)^p \lambda^{p} \}^2K_h(x_i-x)}\\
& \approx \frac{1}{(k+1)!}\lambda^{k+1} g(x)  \frac{\int_a^b (z-x)^{k+1}\{\sum_{p=0}^k \frac{1}{p!}(z-x)^p \lambda^{p}\}f(z)K_h(z-x)dz}{\int_a^b\{\sum_{p=0}^k \frac{1}{p!}(z-x)^p \lambda^{p} \}^2f(z)K_h(z-x)dz}
\end{split}
\end{equation}

We consider the fractional part of the above expression. Let $\frac{z-x}{h}=w$, then the numerator 

\begin{equation*}
\begin{split}
& \int_a^b (z-x)^{k+1}\left\{\sum_{p=0}^k \frac{1}{p!}(z-x)^p \lambda^{p}\right\}f(z)K_h(z-x)dz \\
&= \int (hw)^{k+1}\left\{\sum_{p=0}^k \frac{1}{p!}(hw)^p \lambda^{p}\right\}f(x+hw)K(w)dw \\
&= \int (hw)^{k+1}\left\{\sum_{p=0}^k \frac{1}{p!}(hw)^p \lambda^{p}\right\}(f(x)+hwf'(x))K(w)dw \\
&= \int\left \{ \sum_{p=0}^k\frac{1}{p!}h^{p+k+1}w^{p+k+1} \lambda^{p}f(x)+\sum_{p=0}^k\frac{1}{p!}h^{p+k+2}w^{p+k+2} \lambda^{p}f'(x) \right\}K(w)dw
\end{split}
\end{equation*}

and the denominator

\begin{equation*}
\begin{split}
& \int_a^b\left\{\sum_{p=0}^k \frac{1}{p!}(z-x)^p \lambda^{p}\right \}^2f(z)K_h(z-x)dz \\
& = \int \left\{\sum_{p=0}^k \frac{1}{p!}(hw)^p \lambda^{p} \right\}^2f(x+hw)K(w)dw \\
& \approx f(x)
\end{split}
\end{equation*}

Therefore, when $k$ is an odd number, the asymptotic conditional bias of $\widehat{g}_k (x)$ is

\begin{equation}
\mathrm{Bias}(\widehat{g}_k (x)|x_1,...,x_n) = \frac{1}{(k+1)!}\lambda^{k+1}g(x)h^{k+1}\mu_{k+1}+o_p(h^{k+1})
\end{equation}

where $\mu_{k+1}=\int w^{k+1}K(w)dw < \infty$.

and when $k$ is an even number, 

\begin{equation}
\mathrm{Bias}(\widehat{g}_k (x)|x_1,...,x_n) = \frac{1}{(k+1)!}\lambda^{k+1}g(x)h^{k+2}\mu_{k+2}(\lambda+\frac{f'(x)}{f(x)})+o_p(h^{k+2})
\end{equation}

where $\mu_{k+2}=\int w^{k+2}K(w)dw < \infty$.

\subsection{The Asymptotic Conditional Variance}

The asymptotic conditional variance of the $k^{th}$-degree estimator (DE1-$k$ estimator) is

\begin{align}
\mathrm{Var}(\widehat{g}_k (x)|x_1,...,x_n)
&= \frac{\sum_{i=1}^n  \mathrm{Var}(y_i|x_1,...,x_n) \{\sum_{p=0}^k \frac{1}{p!}(x_i-x)^p \lambda^{p} \}^2 K_h^2(x_i-x) }{ \{\sum_{i=1}^n \{\sum_{p=0}^k \frac{1}{p!}(x_i-x)^p \lambda^{p} \}^2K_h(x_i-x) \}^2} \nonumber \\
&= \frac{\sum_{i=1}^n  \sigma^2 \{\sum_{p=0}^k \frac{1}{p!}(x_i-x)^p \lambda^{p} \}^2 K_h^2(x_i-x) }{ \{\sum_{i=1}^n \{\sum_{p=0}^k \frac{1}{p!}(x_i-x)^p \lambda^{p} \}^2K_h(x_i-x) \}^2} \nonumber \\
& \approx \frac{1}{n} \frac{\int_a^b  \sigma^2 \{1+\sum_{p=1}^k \frac{1}{p!}(z-x)^p \lambda^{p} \}^2 K_h^2(z-x)f(z)dz }{ \{\int_a^b \{1+\sum_{p=1}^k \frac{1}{p!}(z-x)^p \lambda^{p} \}^2K_h(z-x) f(z)dz\}^2} \nonumber \\
&=\frac{1}{nh} \frac{\int  \sigma^2 \{1+\sum_{p=1}^k \frac{1}{p!}(hw)^p \lambda^{p} \}^2 K^2(w)f(x+hw)dw }{ \{\int \{1+\sum_{p=1}^k \frac{1}{p!}(hw)^p \lambda^{p} \}^2K(w) f(x+hw)dw\}^2} \nonumber \\
&\approx \frac{\sigma^2R(K)}{f(x)}+o_p\left(\frac{1}{nh}\right)
\end{align}

where $\frac{z-x}{h}=w$.

The ratio of these expressions is independent of $\sigma^2$.  Further direct analysis
of the ratio is difficult, but simulation provides some insight.  

For a small study, we assume $\lambda = 1$ and we take a sample of size 10 with a 
random uniform sampling design.   We then calculate the ratio of the variances at
each design point using a gaussian kernel.

We find that the average of the ratio values is $0.998$ and
the range of values is 
(0.893, 1.096).  These results are typical.  Thus, we conclude that the
variance behaviour of the first order approximation to the exponential growth
model is quite similar to that of local constant regression.

\section{Simulation Study}

We now study the method numerically by simulating 100 sets of data (with sample sizes 25 and 10) under three scenarios:
\begin{itemize}
\item Scenario 1: $g(x) = \mbox{e}^x$ (pure exponential growth)
\item Scenario 2: $g(x) = \mbox{e}^{x - 0.025x^2}$ (damped exponential growth)
\item Scenario 3: $g(x) = \mbox{e}^{x - 0.1x^2}$ (heavily damped exponential growth)
\end{itemize}
In each case,  noise with standard deviation 0.1 has been added.    

For each simulated dataset, we apply the following estimators: NW (local constant), LL (local linear), LQ (local quadratic), LC (local cubic), DE1-1 (first-degree DE-Assisted exponential growth approximation), 
DE1-2 (second-degree DE-Assisted exponential growth approximation), DE1-3 (third-degree DE-Assisted exponential growth approximation), DE1-4(fourth-degree DE-Assisted exponential growth approximation), DE1-5(fifth-degree DE-Assisted exponential growth approximation), and NLS( the exponential growth model fit by nonlinear least-squares). Bandwidths are selected by leave-one-out cross-validation.

Our basis for comparing the methods is the median absolute deviation (MAD) which is calculated by taking the absolute value of the difference between the fitted values and the corresponding true model values and finding the median. That is, 
$$\mathrm{MAD}= \mathrm{median}(|y_i-\hat{y}_i|)$$
We then report the average of the MADs for the set of 100 simulated datasets.  We also display the distribution of MAD values in boxplots for each method. 

In our local exponential growth model, the governing differential equation is assumed to be $g' = \lambda g$ in each case. Thus, the second and third scenarios represent model misspecification scenarios.
 
To specify the issue of model misspecification, we consider two different model misspecification scenarios with different levels of misspecification. That is, Scenario 3 is misspecified more seriously than Scenario 2. We want to see the estimation behavior of DE-assisted estimators in different situations.

\subsection{Numerical Properties of Estimators Under Uniform Sampling Design}

In our first set of simulations, we consider a random uniformly distributed design for the covariate on the interval $[0, 1]$.  

\begin{figure}
        \centering
        \vspace{-18mm}
        \includegraphics[width=1\textwidth]{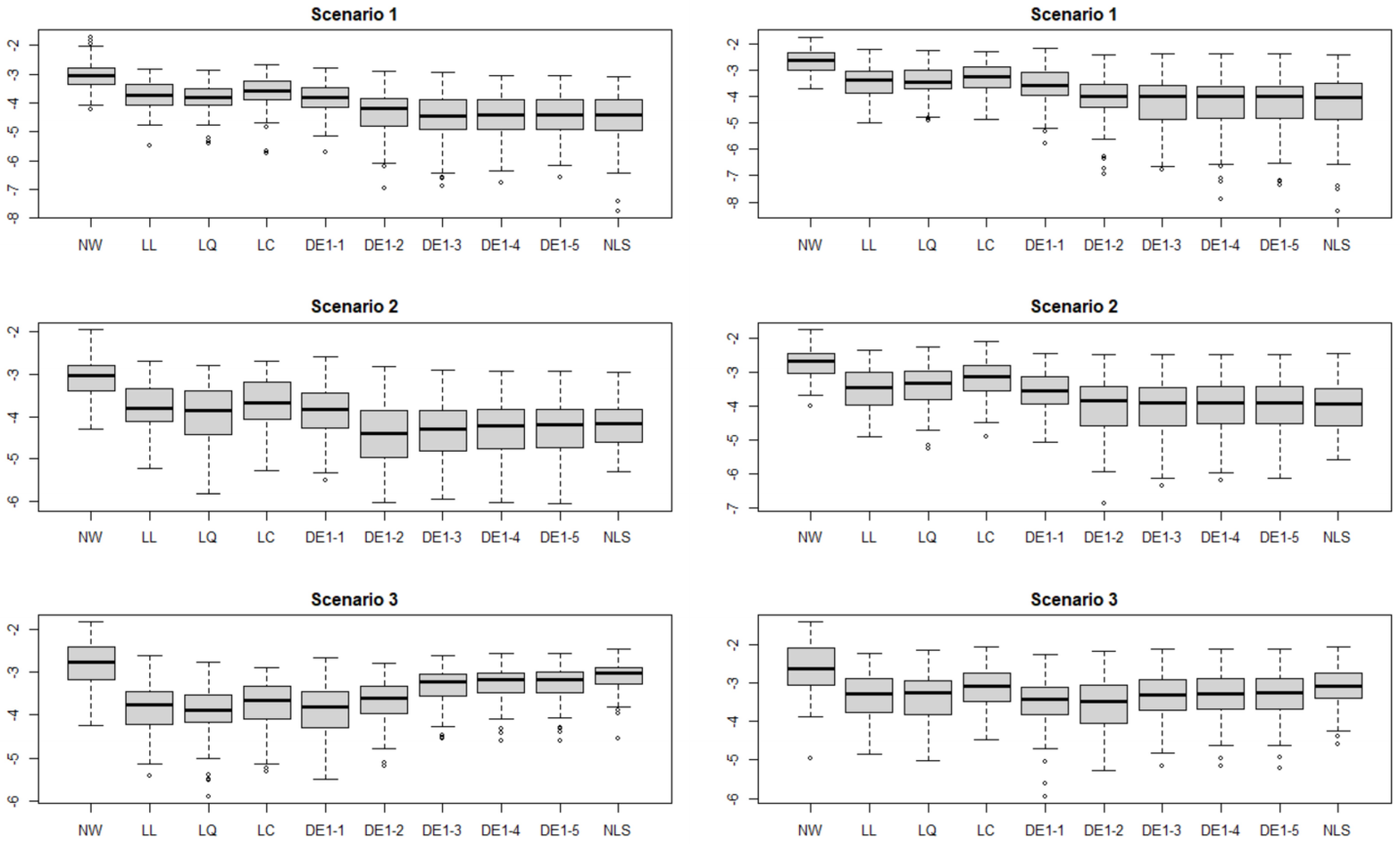}
        \vspace{-20mm}
        \caption{MAD (Median Absolute Deviation) distributions (on the log scale) for the estimation methods
applied to simulated data under the three scenarios in uniform design. Sample size = 25. Note: $\mathrm{MAD}=\mathrm{median}(|y_i-\hat{y}_i|).$}
        \label{fig:unif}
        \end{figure}

Boxplots are displayed in Figure \ref{fig:unif} which shows the distributions (on the log scale) of the Means of Median Absolute Deviation (MAD) for the different estimation methods, NW (local constant), LL (local linear), LQ (local quadratic), LC (local cubic), DE1-1 (first-degree DE-Assisted regression), DE1-2 ( second-degree DE-Assisted regression), and DE1-5 (fifth-degree DE-Assisted regression), applied to simulated data under the three scenarios with a uniform design. Table \ref{table:unifMean} contains the average MAD (Median Absolute Deviation) for each method for the four different scenarios and for the two sample sizes. Table \ref{table:unifError} provides the corresponding standard error estimates.  

 \begin{table}[ht]
\centering
\scalebox{0.9}{
\begin{tabular}{ccccccc}
  \hline
 & Scen. 1 (25) & Scen. 2 (25) & Scen. 3 (25) & Scen. 1 (10) & Scen. 2 (10) & Scen. 3 (10) \\ 
  \hline
  NW & 53.27 & 52.97 & 69.22 & 77.14 & 71.70 & 84.64 \\ 
  LL & 26.19 & 25.92 & 24.86 & 37.88 & 37.22 & 41.06 \\ 
  LQ & 24.32 & 24.29 & 23.20 & 38.59 & 39.08 & 41.44 \\ 
  LC & 29.52 & 28.83 & 26.81 & 43.00 & 47.43 & 49.18 \\ 
  DE1-1 & 24.62 & 24.39 & 24.25 & 35.28 & 33.97 & 35.90 \\ 
  DE1-2 & 15.63 & 15.78 & 27.12 & 23.50 & 24.79 & 34.97 \\ 
  DE1-3 & 14.58 & 15.68 & 38.58 & 21.99 & 24.60 & 40.18 \\ 
  DE1-4 & 14.46 & 16.69 & 40.51 & 21.94 & 24.99 & 41.14 \\ 
  DE1-5 & 14.45 & 16.86 & 40.84 & 21.94 & 25.01 & 41.30 \\ 
  NLS & 14.41 & 17.53 & 47.25 & 21.54 & 24.56 & 49.39 \\ 
  \hline  
\end{tabular}
}
\caption{Means of Median Absolute Deviation (MAD) multiplied by 1000 based on 100 simulation runs for the three scenarios under the uniform sampling design. Sample sizes are specified in the brackets.} 
\label{table:unifMean}
\end{table}

\begin{table}[ht]
\centering
\scalebox{0.9}{
\begin{tabular}{ccccccc}
  \hline
 & Scen. 1 (25) & Scen. 2 (25) & Scen. 3 (25) & Scen. 1 (10) & Scen. 2 (10) & Scen. 3 (10) \\ 
  \hline
  NW & 3.01 & 2.69 & 3.50 & 3.42 & 2.99 & 5.07 \\ 
  LL & 1.19 & 1.40 & 1.36 & 2.18 & 2.22 & 2.27 \\ 
  LQ & 1.11 & 1.38 & 1.24 & 2.19 & 2.13 & 2.43 \\ 
  LC & 1.26 & 1.54 & 1.27 & 2.35 & 2.49 & 2.46 \\ 
  DE1-1 & 1.23 & 1.45 & 1.39 & 2.28 & 1.93 & 1.99 \\ 
  DE1-2 & 0.94 & 1.17 & 1.16 & 1.78 & 1.86 & 2.18 \\ 
  DE1-3 & 0.96 & 1.03 & 1.30 & 1.87 & 1.84 & 2.23 \\ 
  DE1-4 & 0.95 & 1.06 & 1.32 & 1.86 & 1.84 & 2.26 \\ 
  DE1-5 & 0.95 & 1.05 & 1.33 & 1.86 & 1.84 & 2.27 \\ 
  NLS & 0.95 & 0.99 & 1.29 & 1.82 & 1.82 & 2.18 \\ 

  \hline 
\end{tabular}
}
\caption{Standard errors of mean for Median Absolute Deviation (MAD) multiplied by 1000 based on 100 simulation runs for the three scenarios under the uniform sampling design. Sample sizes are specified in the brackets.} 
  \label{table:unifError}            
\end{table}

What we see under the first scenario with a sample of size 25 is that the solution of the differential equation
is the most accurate estimator, followed by the fifth order DE approximation.  The 
accuracy of the methods decreases as the order decreases.  The completely nonparametric
methods are not competitive with the DE-assisted methods in this case.

When there has been some model misspecification as in Scenario 2, the solution of the
DE is no longer the best estimator.  However, the approximate methods still work fairly
well.  The nonparametric methods remain uncompetitive.

In the third scenario, none of the methods are very accurate, but the first order DE-assisted method performs substantially better than any of the methods.  

The results are almost qualitatively the same when the sample size is smaller.  The only
difference is that the exponential growth model remains somewhat more competitive under
Scenario 2.

According to the results of boxplots and tables of the local growth model, we find that, when the model is correct (Scenario 1), the highest-degree estimator (DE1-5) has the lowest bias. But when the model is misspecified (such as Scenario 2, 3), the higher-degree estimators (DE1-3, DE1-4, and DE1-5) have higher biases than the lower-degree estimator(DE1-1 and DE1-2). For the model selection, a higher-degree DE-Assisted regression is preferable. However, when the model is wrong, the higher-degree estimator will have greater bias. A first-degree or second-degree estimator will be a better choice.

\subsection{Numerical Properties of Estimators Under Sparse Design}

In our second set of simulations, we consider a random beta distributed design for the covariate on the interval $[0, 1]$.  Shape parameters for the beta distribution were taken to be 1.0 and 0.5.  This
induces sparsity near the origin.  

\begin{figure}
        \centering
        \vspace{-18mm}
        \includegraphics[width=1\textwidth]{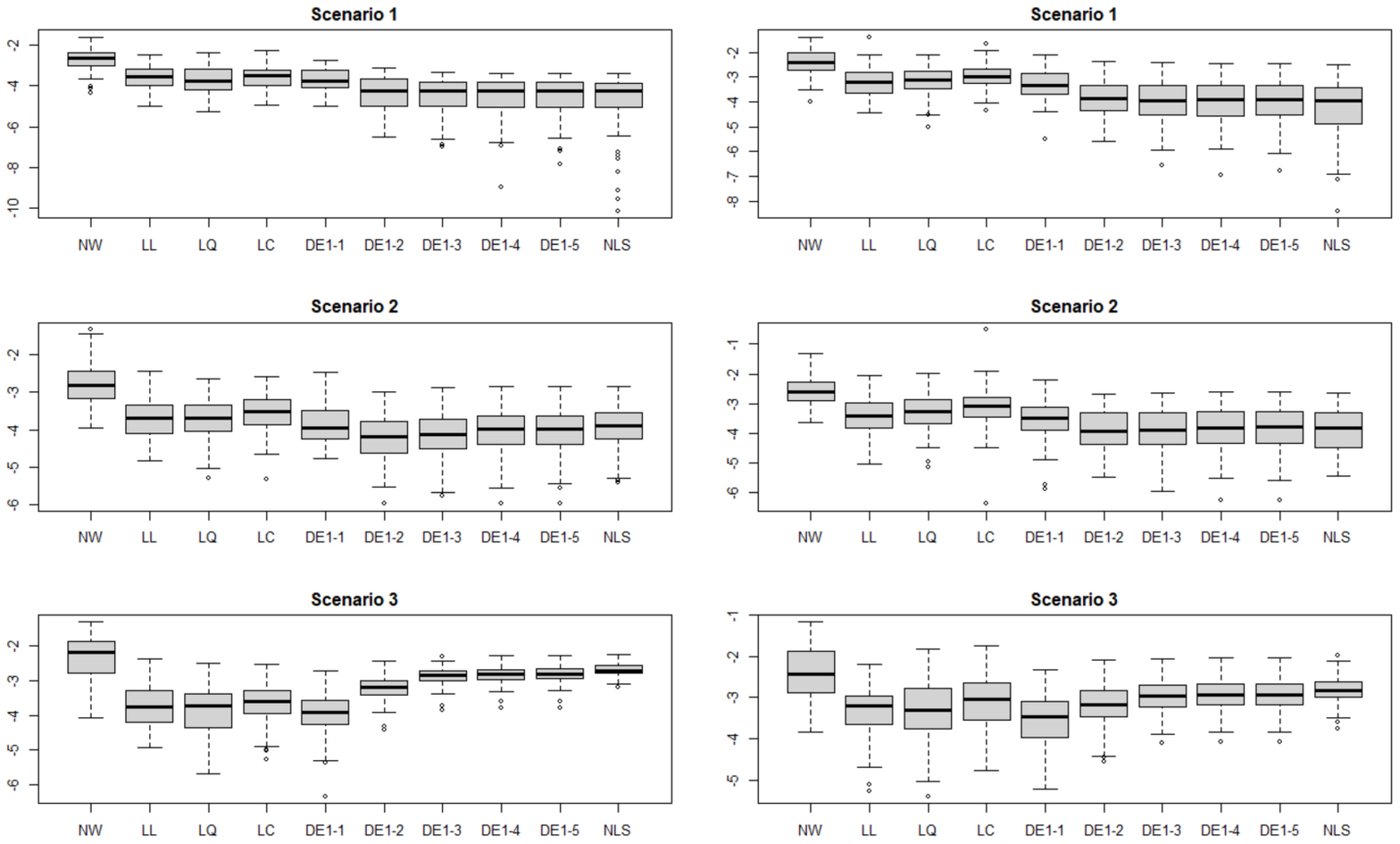}
        \vspace{-20mm}
        \caption{MAD (Median Absolute Deviation) distributions (on the log scale) for the estimation methods
applied to simulated data under the three scenarios in sparse design. Sample size = 25. Note: $\mathrm{MAD}=\mathrm{median}(|y_i-\hat{y}_i|).$}
        \label{fig:spa}
        \end{figure}

Boxplots of the MAD distributions for the various methods applied to the sparse designs are displayed in Figure \ref{fig:spa}.  Tables \ref{table:spaMean} and \ref{table:spaError} provide analogous information
to that of Tables \ref{table:unifMean} and \ref{table:unifError}.  

\begin{table}[ht]
\centering
\scalebox{0.9}{
\begin{tabular}{ccccccc}
  \hline
 & Scen. 1 (25) & Scen. 2 (25) & Scen. 3 (25) & Scen. 1 (10) & Scen. 2 (10) & Scen. 3 (10) \\ 
  \hline
  NW & 72.17 & 71.83 & 115.81 & 101.71 & 85.90 & 110.39 \\ 
  LL & 32.28 & 28.84 & 28.25 & 47.91 & 39.11 & 40.66 \\ 
  LQ & 27.90 & 27.92 & 25.59 & 47.92 & 42.20 & 46.23 \\ 
  LC & 31.87 & 32.07 & 29.84 & 58.45 & 55.30 & 53.53 \\ 
  DE1-1 & 28.72 & 25.13 & 23.41 & 43.27 & 34.24 & 33.95 \\ 
  DE1-2 & 16.62 & 17.20 & 41.93 & 26.07 & 24.52 & 46.52 \\ 
  DE1-3 & 14.50 & 18.40 & 57.69 & 23.79 & 24.94 & 54.60 \\ 
  DE1-4 & 14.20 & 20.12 & 60.54 & 23.61 & 25.76 & 56.05 \\ 
  DE1-5 & 14.19 & 20.40 & 60.99 & 23.61 & 25.87 & 56.24 \\ 
  NLS & 13.97 & 21.74 & 67.59 & 21.39 & 24.89 & 64.62 \\ 
  \hline  
\end{tabular}
}
\caption{Means of Median Absolute Deviation (MAD) multiplied by 1000 based on 100 simulation runs under the three sparse design scenarios. Sample sizes are specified in the brackets.} 
\label{table:spaMean}
\end{table}

\begin{table}[ht]
\centering
\scalebox{0.9}{
\begin{tabular}{ccccccc}
  \hline
 & Scen. 1 (25) & Scen. 2 (25) & Scen. 3 (25) & Scen. 1 (10) & Scen. 2 (10) & Scen. 3 (10) \\ 
  \hline
  NW & 3.26 & 4.61 & 6.29 & 5.10 & 4.75 & 6.92 \\ 
  LL & 1.65 & 1.62 & 1.70 & 3.19 & 2.30 & 2.18 \\ 
  LQ & 1.74 & 1.43 & 1.68 & 2.49 & 2.08 & 3.07 \\ 
  LC & 1.65 & 1.39 & 1.53 & 3.16 & 6.11 & 3.24 \\ 
  DE1-1 & 1.50 & 1.52 & 1.40 & 2.57 & 2.09 & 1.95 \\ 
  DE1-2 & 1.10 & 1.02 & 1.30 & 1.75 & 1.54 & 2.31 \\ 
  DE1-3 & 0.91 & 1.07 & 1.32 & 1.73 & 1.60 & 2.24 \\ 
  DE1-4 & 0.90 & 1.12 & 1.36 & 1.74 & 1.63 & 2.25 \\ 
  DE1-5 & 0.90 & 1.12 & 1.37 & 1.74 & 1.62 & 2.25 \\ 
  NLS & 0.93 & 1.09 & 1.23 & 1.70 & 1.61 & 2.12 \\ 
  \hline  
\end{tabular}
}
\caption{Standard errors of mean for Median Absolute Deviation (MAD) multiplied by 1000 based on 100 simulation runs under the three sparse design scenarios. Sample sizes are specified in the brackets.} 
\label{table:spaError}
\end{table}

The results are similar to those reported for the uniform design.  The completely nonparametric methods are not competitive with the DE-assisted methods, even under model misspecification. 

\section{Comparison of Estimates}

We will compare our DE-assisted estimators with the traditional local polynomial regression estimators and \citet{he2009double}'s double-smoothing estimator. 

In this section, we will compare the asymptotic properties for different estimates both for correct model and misspecified model.

When the DE-Assisted local exponential growth model (\ref{equ:exp}) we use is correct, which means that the truth is $g(x)=e^{\lambda_1x}$, and then $g^{(k)}=\lambda^k g(x)$, where $k=1,2,...$. Table \ref{table:asymptotic} gives us a summarized table with comparing different estimators. The order follows the pattern from the largest to the smallest approximately. 

The summary table of bias and variance shows that, for the DE-Assisted local growth model, the higher-degree estimator has a lower bias while the variance stays the same. So if the model is correct and we are not concerned about the computational load, we prefer a higher-degree estimator. However, if the model is misspecified, it will be a different story. The lower-degree estimator is better than higher-degree estimator. Sometimes when the model is badly misspecified, the first-degree estimator is probably the best choice. We will demonstrate this result in the simulation study.

\begin{table}[h!]
  \begin{center}  
    \begin{tabular}{c|c|c} 
      \hline
      \textbf{Method} & \textbf{Asymptotic bias in interior} & \textbf{Asymptotic variance}\\
      \hline
      NW & $\frac{1}{2}\{\lambda^2g(x)+2\lambda\frac{g(x)f'(x)}{f(x)}\}h^2 \mu_2+o_p(h^2)$  & $\frac{1}{nh}\frac{\sigma^2}{f(x)}v_0+o_p(\frac{1}{nh})$ \\
      LL & $\frac{1}{2}\lambda^2g(x)h^2 \mu_2 +o_p(h^2)$ & $\frac{1}{nh}\frac{\sigma^2(x)}{f(x)}v_0+o_p(\frac{1}{nh})$\\
      DE1-1 & $\frac{1}{2}\lambda^2g(x)h^2\mu_2+o_p(h^2)$ & $\frac{1}{nh}\frac{\sigma^2}{f(x)}v_0+o_p(\frac{1}{nh})$ \\
      DE1-2  & $\frac{1}{6}\lambda^3g(x)h^4\mu_4(\lambda + \frac{f'(x)}{f(x)})\mu_4+ o_p(h^4)$ & $\frac{1}{nh}\frac{\sigma^2}{f(x)}v_0+o_p(\frac{1}{nh})$\\  
      LQ  & $\frac{1}{24}\frac{\mu_2\mu_6-\mu_4^2}{\mu_2^2-\mu_4}\{\lambda^4g^(x)+4\lambda^3\frac{g(x)f'(x)}{f(x)}\}h^4+o_p(h^4)$  & $\frac{1}{nh}\frac{\sigma^2}{f(x)}\frac{\mu_4^2v_0-2\mu_2\mu_4v_2+\mu_2^2v_4}{(\mu_2^2-\mu_4)^2}+o_p(\frac{1}{nh})$\\ 
      LC & $ \frac{1}{24}\frac{\mu_2\mu_6-\mu_4^2}{\mu_2^2-\mu_4}\lambda^4g(x)h^4+o_p(h^4)$  & $\frac{1}{nh}\frac{\sigma^2}{f(x)}\frac{\mu_4^2v_0-2\mu_2\mu_4v_2+\mu_2^2v_4}{(\mu_2^2-\mu_4)^2}+o_p(\frac{1}{nh})$\\ 
      DS & $h^4B(x)+o_p(h^4)$ & $\frac{1}{nh}\frac{\sigma^2}{f(x)}V+o_p(\frac{1}{nh})$\\       
      DE1-3 &$\frac{1}{24}\lambda^4g(x)h^4\mu_4+o_p(h^4)$ & $\frac{1}{nh}\frac{\sigma^2}{f(x)}v_0+o_p(\frac{1}{nh})$\\
     \hline      
    \end{tabular}    
    \caption{Summary of asymptotic conditional bias and variance for the NW (local constant) , LL (local linear) , LC (local cubic), DS (double smoothing), DE1-1 (first-degree DE-Assisted), DE1-2 (second-degree DE-Assisted), and DE1-3 (third-degree DE-Assisted) estimators for correct models.\\
Note:  In the DS estimator, $B(x)=(\mu_2^2-\mu_4)/4[g''(x)f''(x)/f(x)+2(g^{(3)}(x)f'(x)/f(x))+g^{(4)}(x)] = (\mu_2^2-\mu_4)/4[\lambda^2g(x)f''(x)/f(x)+2(\lambda^3g(x)f'(x)/f(x))+\lambda^4g(x)] $.  $V=\int \{(K*K)(v)-(K_1*K_1)(v)/\mu_2\}^2dv$, where $K_1(u)=uK(u)$.  $f(x)$ is the design density which is the density of the covariate $X$, $\mu_k = \int w^k K(w)dw$ and $v_k = \int w^kK^2(w)dw$, $k=0,1,2,...$. The detail of the DS estimator, which used a double-smoothing technique, can be found in the paper \citet{he2009double}. }
\label{table:asymptotic} 
  \end{center}   
\end{table}

According to the result of simulation study, we realize that, in a misspecified model, a higher-degress DE-Assisted estimator is not a better choice than a lower-degree one. When the model is badly misspecified, a first-degree estimator is competitive enough among the regression estimators. Therefore, we will compare DE1-1 estimate with local constant and local linear estimates for misspecified models as follows.

The model we use is the DE-Assisted local exponential growth model (\ref{equ:exp}) with $g'(x)=\lambda_1 g(x)$.

The truth is the scenario,  $g(x)=e^{\lambda_1 x - \lambda_2 x^2}$, where $\lambda_1$ and $\lambda_2$ are known and global parameters, $\lambda_1 >0$ and $\lambda_2>0$. Therefore, the truth is, 

$$g'(x)=(\lambda_1-2x\lambda_2)g(x),$$

and $$g''(x)=((\lambda_1 - 2x\lambda_2)^2 - 2x \lambda_2)g(x).$$

Using the previous technique we applied in Section 3.3, we obtain the asymptotic conditional biases of DE1-1, NW (local constant), and LL (local linear) estimators. Table \ref{table:asymptoticmis} does not include the asymptotic conditional variances since they have no changes for misspecified models. We can find the DE1-1 estimator has the smallest bias among thess three ones in Table \ref{table:asymptoticmis}.

\begin{table}[h!]
  \begin{center}  
    \begin{tabular}{c|c}
      \hline
      \textbf{Method} & \textbf{Asymptotic bias in interior}\\
      \hline
      NW & $\frac{1}{2}((\lambda_1 - 2x\lambda_2)^2 - 2x \lambda_2)g(x)h^2 \mu_2+\frac{g'(x)f'(x)}{f(x)}h^2 \mu_2+o_p(h^2)$ \\
      LL & $\frac{1}{2}((\lambda_1 - 2x\lambda_2)^2 - 2x \lambda_2)g(x)h^2 \mu_2+o_p(h^2)$  \\
      DE1-1 & $\frac{1}{2}((\lambda_1 - 2x\lambda_2)^2 - 2x \lambda_2)g(x)h^2 \mu_2 - 2x\lambda_2 g(x)(\lambda_1+\frac{f'(x)}{f(x)})h^2 \mu_2+o_p(h^2)$  \\
      \hline      
    \end{tabular}    
    \caption{Summary of asymptotic conditional bias and variance for the NW (local constant), LL (local linear), and DE1-1 (first-degree DE-Assisted) estimators for misspecified models.} 
    \label{table:asymptoticmis} 
  \end{center}   
\end{table}

In the next section, we will extend local growth model and apply it to the mouse tumor volume data which we described in Section 1.

 \section{Local Sub-Exponential Growth Model and Application}

\subsection{Local Sub-Exponential Growth Model}

The exponential model often leads to overestimates of growth at later times.
Therefore, modifications such as the Gompertz model have been considered.  
Another possibility is as follows.   With $\alpha \in (0, 1)$, suppose
\begin{equation}\label{eqn:DEgrowth}  g' = \lambda g^\alpha \end{equation}
The explicit solution to this differential equation is 
\begin{equation}\label{eqn:soln} g(x) = \{(1-\alpha) (\lambda x + g(0))\}^{1/(1-\alpha)}. \end{equation}

Alternatively, we can apply our proposed method to obtain a  local solution of the differential equation by 
constructing the Taylor expansion about an evaluation point $x_0$, 
and applying the differential equation:
\[ g'(x_0) = \lambda (g(x_0))^\alpha \]
and the second derivative:
\[ g''(x_0) = \alpha \lambda^2 (g(x_0))^{2\alpha - 1}.  \]
Upon defining
\[ g^*(x_i) = g(x_0) + (x_i-x_0) \lambda (g(x))^\alpha + \frac{(x_i-x_0)^2}{2} \alpha \lambda^2 (g(x))^{2\alpha - 1}\]
our DE-assisted local quadratic regression estimator at $x_0$ is obtained by 
minimizing
\[ \sum_{i=1}^n (y_i - g^*(x_i))^2 K_h(x_i-x_0) \]
with respect to the single parameter $g(x_0)$.  The minimization is obtained
by solving a weighted nonlinear least-squares problem which
is easily solved iteratively, given an appropriate initial guess.    One possibility is to simply
use the local constant regression estimate as the starting value for the iteration.

\subsection{Application to Sparse Tumor Growth Data}

 Since the growth is roughly approximated by an exponential model, it makes more sense
 to log transform the response variable and to make the assumption that the errors are
 additive on the log scale.  Under a normal distribution assumption, this amounts to
 the assumption that, on the original scale, the errors are multiplicative and distributed according
 to a log-normal distribution.  
 
 After transformation, the differential equation model at (\ref{eqn:DEgrowth}) becomes
\begin{equation}\label{eqn:logGrowth}  G'(x) = \lambda \mbox{e}^{(\alpha - 1) G(x)} \end{equation}
where $G(x) = \log(g(x))$.   The model on the data becomes
\begin{equation}\label{eqn:0} \log(y_i)  = G(x_i) + \varepsilon_i \end{equation}
and first order and second order Taylor expansions about $x_0$ give
\begin{equation}\label{eqn:1st} \log(y_i) \doteq G(x_0) + \lambda \mbox{e}^{(\alpha - 1)G(x_0)}(x_i -  x_0) + \varepsilon_i \end{equation}
and 

\begin{equation}\label{eqn:2nd} \log(y_i) \doteq G(x_0) + \lambda \mbox{e}^{(\alpha - 1)G(x_0)}(x_i - x_0) +
\frac{1}{2}\lambda^2(\alpha - 1)\mbox{e}^{2(\alpha - 1) G(x_0)}(x_i - x_0)^2 + 
 \varepsilon_i \end{equation}
When implemented in the DE-assisted regression methodology, we refer to the models
(\ref{eqn:1st}) and (\ref{eqn:2nd}) as the first and second order local growth models, respectively.

Since we do not know the true model for this data set, we arbitrarily choose the local
linear estimate for the full data set as the ``truth''.  The bandwidth  is obtained using 
the \verb!dpill! function \citep{ruppert1995effective} in the \textit{KernSmooth} package \citep{kernsmooth} in R:  $h = 2.38$.  The standard
deviation of the residuals is obtained as 0.089.     The local linear fit $\widehat{G}_{\ell \ell}(x)$, together with
the standard deviation become the basis of another simulation study whereby new
observations are generated at the original design points $x_i$ according to a normal distribution with mean $\widehat{G}_{ll}(x_i)$ ($i \in 1, 2, \ldots, 10$) and the empirical standard deviation.  

We train our competing models on the simulated datasets with observations 4 through 8 removed
each time.    The models under test are local constant, local linear, local quadratic,
first order local growth, second order local growth, and the nonlinear least-squares estimate
of the solution to the differential equation (\ref{eqn:logGrowth}).  

Estimation of $\alpha$ and $\lambda$ is needed for the two local growth models.   
From (\ref{eqn:soln}), we can see that 
\[ G(x) \doteq \frac{1}{1-\alpha} (\log(1-\alpha) + \log(\lambda) + \log(x)) \]
since $g(0)$ is necessarily a very small value in this application.    This means
that the simple linear regression slope estimator for the model $\log(y)$ versus $\log(x)$ is an
estimator for $1/(1-\alpha)$.  We use this to estimate $\alpha$.  Given this estimate,
we then estimate $\lambda$ by applying nonlinear least-squares to the model
\[ y = \{(1-\widehat{\alpha}) (\lambda x)\}^{1/(1-\widehat{\alpha})}. \]
This is, once more, based on an approximation to the explicit DE solution given at (\ref{eqn:soln}).  

The squared differences between $\widehat{G}(x_i)$ and $\widehat{G}_{\ell \ell}(x_i)$ are
calculated for $i = 4, \ldots, 8$ and averaged, for each of the six estimation methods.  The
averages of these average squared differences are listed in 
Table \ref{logmodel}.  The squared differences were calculated, both on the raw
scale, and on the log scale.   The smallest values in both cases are for the second
order local growth model, with the first order local growth model close behind.  The
local linear model also enjoys fairly small average squared errors, but there is a slight
bias in favour of local linear, since the underlying data follows a linear model.
Local quadratic regression performs somewhat worse, and the nonlinear growth model
has considerably larger errors than the lower order local approximations.   This suggests
that the suggested local growth model may not really be appropriate for the data.
However, the overall message is that the DE model can guide the kernel regression
methods to a satisfactory estimate.

\begin{table}[ht]
\centering
\begin{tabular}{ccc}
  \hline
 & log scale & original scale \\ 
  \hline
  NW & 0.48 & 0.44 \\ 
  LL & 0.03 & 0.03 \\ 
  LQ & 0.05 & 0.06 \\ 
  DE1-1 & 0.03 & 0.02 \\ 
  DE1-2 & 0.03 & 0.03 \\ 
  NLS & 0.79 & 0.09 \\ 
   \hline
\end{tabular}
\caption{Average squared error summaries for each of the modelling approaches, NW (local constant), LL ( local linear), LQ (local quadratic), DE1-1 (first-degreee DE-Assisted regression), DE1-2 (second-degree DE-Assisted regression), and NLS (nonlinear least square). Errors in the first column are based on differences between fitted and observed values on the log scale, and errors in the second column are based on differences between exponentiated fitted values and raw observed values.} 
\label{logmodel}
\end{table}

 \section{Discussion}

Differential equation information can be used to improve local polynomial regression estimates.
The proposed method is simple and without requiring substantial amounts of parameter tuning.
The solution of the DE is not needed, and the method is only slightly more
complicated to implement than local constant regression.  Thus, it is simple and
the computation is fast.

The number of derivatives to evaluate will depend on how believable the DE is. 
If the DE is not very believable, use fewer terms and a smaller $h$
Weighted nonlinear least-squares may be needed to do the estimation, if the 
DE is not linear.
Global parameters (e.g. $\lambda, \alpha$ in growth model) need to be estimated. 

Bias can be reduced, relative to local constant regression, both in the interior and at the data boundaries.    The method
appears to be somewhat robust to model misspecification, and it seems to work
particularly well, as compared with other methods, when the design is sparse.  

Bias reduction can still be achieved even if the parameters
are estimated.   Bias order remains $O(h^2)$ as long as parameter estimates
converge at rate $n^{1/2}$ (typical parametric rate) 
and even if parameter estimates converge at rate $n^{2/5}$ 
(the typical nonparametric rate).  

We have focussed on a simple case, the exponential model, as a proof of concept. In the future, study of more complicated models will be of interest.

{\bf Acknowledgements} 

This research has been supported in part by a grant from the Natural Sciences and
Engineering Research Council of Canada.

\end{document}